# Optimization of a Microsphere-based Variable Density Multilayer Insulation System for Cryogenic Applications

**Ritayan Mukherjee**

f20212504@goa.bits-pilani.ac.in

Birla Institute of Technology and Science Pilani, K. K. Birla Goa Campus, Goa, India

**Arun Srinivasan**

aruns@nal.res.in

Council of Scientific and Industrial Research, National Aerospace Laboratories, Bengaluru, India

**Abstract**

One of the standard composite insulation systems for cryogenic applications consists of a layer of hollow glass microspheres (HGMs) followed by a layer of variable density multilayer insulation (VDMLI) comprised of various internal combinations of reflectors and spacers. Microstructural effects of the HGM assembly and convection between the vessel and its surroundings are unaccounted for in existing calculations, which the current study incorporates in the unit cell-based HGM analytical model. Building on it, the insulation performance of the HGM layer with increasing thickness is studied for a Dewar insulation system containing liquid nitrogen (boiling point 78 K) based on pressures ranging from high vacuum (0.0133 Pa) to atmospheric ($10^5$ Pa) for two benchmark cryogenic vessel sizes. A suitable range of thickness for both systems is suggested. Further, the performance of four arithmetic progression-based VDMLI profiles in combination with an optimal HGM layer is studied. A VDMLI configuration with repeating spacer layers 1/1/2/2/… is found to perform best in high-pressure conditions. Finally, based on heat transfer performance within the given range of pressures and warm boundary temperatures up to 393 K, optimal thickness values of 120-130mm for the HGM layer and 125mm for the best VDMLI configuration are reported.

# 1. Introduction

The applications of cryogenic technology are far-reaching in today's world. For instance, widely available liquid nitrogen (boiling point 78 K) is needed for food processing [1-3], and industrial storage of liquefied natural gas (LNG) involves cryogenic tanks [1, 4, 10]. On a more specialized scale, particle accelerators at CERN have relied on liquid helium (boiling point 4.2 K) for the cooling of superconducting magnets (generally operating at 4.5 K and 1.9 K [5]) in ultra-high-vacuum conditions [6]. Supercooled liquid hydrogen (boiling point 20 K), on the other hand, has gained popularity due to its sustainability [7] and its highly efficient performance as a rocket fuel since its first run in the Atlas-Centaur rocket [8, 9]. Connecting these applications is the need for efficient storage systems for cryo-cooled liquids, since maintaining them below their extremely low boiling points is crucial [10]. This prompts the research of highly efficient insulation systems for cryogenic tanks.

Some of the most common cryogenic insulations comprise of either ultralight materials or multilayer insulations (MLIs) containing reflector sheets and separators (or spacers) [11, 12, 28]. Among the former, aerogel, perlite and glass bubbles have been used to make evacuated insulation layers. MLIs with organic materials as reflectors (aluminized mylar, teflon, etc.) and dacron nets as spacers have been tested [11]. They have been used in NASA's Orion spacecraft to improve insulation performance [12] and CERN's particle accelerators. Lightness of the insulation is a desirable trait due to the application of cryogenics especially in aerial and space vehicles [13]. To optimize the insulation system performance (quantified by low thermal conductivity and heat flux) along with its weight, composite systems have been studied [13, 26]. This paper focuses on a recently developing composite insulation system: hollow glass microspheres (HGMs) with a variable density multilayer insulation (VDMLI).

Allen et al. [14] studied the effectiveness of 3M K1 glass bubble insulation and reported its superior vacuum retention capability, structural stability and insulation performance compared to perlite. They also stated that it performs better than MLI above approximately 4 Pa. Additionally, the potential of HGM-based foam as a flame-resistant insulation is being explored [15, 16]. On the other hand, the performance of a variation of the regular MLI known as the variable density MLI (VDMLI), is under focus in current research because it allows for more optimized use of reflectors and spacers [33]. MLIs of uniform density have been tested extensively for use in the space sector [27]. Among composite insulation systems, different VDMLIs in combination with other known insulations such as spray-on foam and HGMs have been studied and compared [17, 18, 26]. Tested with a liquid nitrogen tank, the effectiveness of a 50-layers-thick VDMLI with 3:2:1 spacer ratio is more than MLI by around 17.5% in terms of heat flux when used in combination with spray-on foam [18].

Xu et al. [19] tested an insulation system with HGMs and vapor-cooled shield (VCS) in combination with a MLI. With liquid hydrogen, they found that the reduced heat leakage by 45% and 81% in high and low vacuum respectively compared to the traditional MLI. Their later study on the same system [20] compared heat leakage through the same system for liquid methane, oxygen and hydrogen, and reported the reduction of heat flux by 26%, 30% and 65% compared to spray-on-foam insulation. P Wang et al. [26] suggested based on analytical calculations that a 3M HGMs-VDMLI composite system is 12% more efficient in preventing heat leakage than a thicker foam-VDMLI system while being around 10% lighter. Additionally, optimization algorithms for VDMLI reflector-spacer distributions have also been suggested based on the insulation system design [21, 26]. Based on an exhaustive total number of spacers and calculation of heat flux iteratively, B Wang et al. [21] used an algorithm to optimize the VDMLI arrangement, which resulted in a 45.5% reduced heat flux compared to uniform density MLI. The reported arrangements had more reflectors and less spacers on the hotter side. The study by P Wang et al. [26] suggested an algorithm based on similar constraints for a HGMs-VDMLI insulation.

However, in previous studies, heat flux calculations have been performed across uniform heat transfer surface areas, and they do not consider convection between the vessel and its environment – namely the cryogen and ambient air. Further, it is desirable to have an analytical model to calculate the conductivity of each composite insulation layer placed on a cylindrical storage tank, alternatively called Dewar systems [22, 23], and standardization of the optimal combined thickness across vessel sizes which can overcome pressure leaks. In the current work, the performance of a cylindrical Dewar insulation system was analysed under varying pressure conditions inside the insulation chamber consisting of a layer of HGMs followed by a VDMLI. Optimization of thickness is done for the HGM layer first for two benchmark diameters, following which the performance of VDMLI arrangements in combination with optimized HGM layers is analysed. The VDMLI profiles considered are arithmetic progression (AP)-based and of two separate types: with consecutive once-repeating layers (like a step function) and without repeating layers (like a linear monotonically increasing function). Based on these results, an optimal thickness of each layer along with the most suitable VDMLI arrangement is suggested.

## 2. Analytical Models

### 2.1. Hollow Glass Microsphere (HGM) layer

The analytical model referred to predict the effective thermal conductivity of the HGM layer subject to varying vacuum pressure conditions is based on that used in Peeketi et al. [24]. Thermal conductivity is affected by the density and distribution of particles at the microsphere level [25]. In contrast to the traditional method that has been used for optimization studies for HGMs [26], this model accounts for microstructural effects like the packing fraction, coordination numbers and external compressive pressure which also affect thermal conductivity of the HGM assembly (eq. (2.1)).

$$k_{eff} = \frac{\eta(N_o C_o^e + N_g C_g^e)}{\pi D} + k_r \qquad (2.1)$$

$$k_r = \frac{8DST^3}{\frac{2}{\epsilon_r} - 0.264} \qquad (2.2)$$

The conductivity of a stacked microsphere assembly was analysed on a unit cell basis. An illustration with the required dimensions is shown in Fig. 1. The conductivity of one unit cell is the same as that of the entire assembly, irrespective of its thickness. In this work, an initial packing efficiency of 68% (body-centred unit cell, close to the random closed packing efficiencies analysed in [24]) and the properties of the standard 3M K1 HGMs are considered for analysis. The HGMs are assumed to be non-overlapping (which sets the parameter $C_o^e$ to zero). For radiative conductivity $k_r$ (eq. (2.2)), an emissivity of 0.5 is assumed. The maximum spacing $z$ (see Fig. 1) between two particles is the distance between the surfaces of two particles along the edge of the unit cell, as in eq. (2.3). It is used in the calculations of dimensionless gap conductance (eq. (2.4)) and consequently the gap-type contact conductance $C_g^c$ (eq. (2.6)). This and the microsphere conductance (eq. (2.5)) account for the effective gap conductance (eq. (2.7)). The fitting parameter ($\zeta$) is taken as 0.625. The Smoluchowski Effect is considered in eq. (2.8) and eq.(2.9) [24]. Qualitatively, it means that the conductivity of the residual gas falls with reduction in pressure in confined spaces (i.e. Knudsen number ($K_n$), the ratio between mean free path of the gas and the gap dimension, approaches 1). The accommodation coefficient $a_c$ between stainless steel and nitrogen is taken as 0.88.

$$z = a - 2R \tag{2.3}$$

$$\xi = ln\left(1 + \frac{\zeta^2 R}{z}\right) \tag{2.4}$$

$$C_s = \frac{\pi K_s (\zeta R)^2}{R} \tag{2.5}$$

$$C_g^c = \pi K_f^c R \xi \tag{2.6}$$

$$C_g^e = \frac{1}{\frac{1}{2C_s} + \frac{1}{C_g^c}} \tag{2.7}$$

$$K_f^c = \frac{K_f}{1 + 2\gamma K_n} \tag{2.8}$$

$$\gamma = \frac{19}{12}\frac{2 - a_c}{a_c} \tag{2.9}$$

Kinematic viscosity $\nu$ (eq. (2.10)) and bulk conductivity (eq. (2.11)) of nitrogen $K_f$ are approximated as in [32] with mean temperature of the warm and cold boundaries of the HGM layer, and are used in the calculation of $K_n$ (eq. (2.12)) and $K_f^c$ (eq. (2.8)).

$$\nu = \frac{1.9836 \times 10^{-6} + 6.525 \times 10^{-8} T}{1.19} \tag{2.10}$$

$$K_f = 3.4 \times 10^{-3} + 8.4732 \times 10^{-5} T + 5.0287 \times 10^{-8} T^2 \tag{2.11}$$

$$K_n = \frac{\lambda}{L_g} \tag{2.12}$$

$$\lambda = \frac{\nu}{P}\sqrt{\frac{\pi R_g T}{2m_f}} \tag{2.13}$$

$$L_g = z + R(1 - \cos(\arcsin(\zeta))) \tag{2.14}$$

A compressive pressure $\sigma_{zz}$ of 1 MPa and a modulus of elasticity $E$ of 1 GPa for quartz glass microspheres are used to compute the final packing fraction $\eta$ and the overlap and gap coordination numbers ($N_o$ and $N_g$), according to the correlations (2.15)-(2.17) proposed in [24].

$$\eta = 1.16 \times (\sigma_{zz}/E)^{0.6}/\eta_0 + \eta_0 \tag{2.15}$$

$$N_o = \eta_0^{1.2}(13.39 \times (\sigma_{zz}/E)^{0.03} - 0.1093) \tag{2.16}$$

$$N_g = (25.16\,\eta_0 - 5.28) - N_o \tag{2.17}$$

## 2.2. Variable Density Multilayer Insulation (VDMLI) Layer

To simulate the effect of uniform or variable-density multilayer insulation, the layer-by-layer model is considered. In the layer $i$ corresponding to the $i^{th}$ reflector starting from the colder boundary, heat transfer occurs through the spacers (solid conduction), the residual gas or air (gas conduction), and through radiation. Conventionally, for each of these modes the heat transfer coefficient is defined as the corresponding conductance per unit area.

The radiation heat transfer coefficient, as in [26], is written as:

$$K_{r,i} = \frac{\sigma(T_h - T_c)(T_h^2 + T_c^2)}{\frac{1}{\epsilon_h} + \frac{1}{\epsilon_c} - 1} \tag{2.18}$$

Here, both emissivity values are taken as 0.5.

The solid heat transfer coefficient, as in the OVDMLI model in [33], is written as:

$$K_{s,i} = \frac{(5.79 \times 10^{-4}) \times T_{mean} \times P_c^{0.45}}{N_c} \tag{2.19}$$

Here, $T_{mean}$ is the average temperature between the warm and cold boundaries; $P_c$ is the gauge compressive pressure, given by $P_c = (2.16 \times 10^{-9}) \times N_{den}^{5.844}$ ($N_{den}$ is the layer density in layers/cm); $N_c$ is the number of interfaces per spacer, which is 2.

The gas conduction heat transfer coefficient, as in [33], is written as:

$$K_{g,i} = \frac{\gamma + 1}{\gamma - 1}\sqrt{\frac{R_g}{8\pi M T_{mean}}} Pa_c(T_h - T_c) \tag{2.20}$$

Here, $\gamma$ is the effective ratio of specific heats (at constant pressure and constant volume) and M is the effective molecular weight of the gas inside the VDMLI chamber. The composition is taken as 30% nitrogen and 70% hydrogen. Accordingly, the mean accommodation coefficient $a_c$ for nitrogen and hydrogen is 0.625. The gas pressure $P$ is a crucial parameter, as it accounts for the effectiveness of the VDMLI insulation along with the boundary temperatures $T_h$ and $T_c$.

The resistance per unit area $R_i$ and effective heat transfer coefficient $K_{eff}$ are found thus:

$$K_i = K_{r,i} + K_{s,i} + K_{g,i} \tag{2.21}$$

$$R_i = 1/K_i \tag{2.22}$$

$$K_{eff} = \frac{1}{\sum_{i=1}^{N} 1/K_i} \tag{2.23}$$

In this work, the input parameters are the VDMLI sequence, the desired thickness of the multilayer insulation and the boundary temperatures. An algorithm was used to iteratively find the value of the effective heat transfer coefficient for the entire VDMLI layer. It involves the following steps.

1. As only the boundary temperatures of the entire VDMLI are known, the temperature profile of the VDMLI is assumed. In the code, it is assumed as per eq. (2.26). This will yield the initial heat transfer coefficient of each layer.
2. The corrected temperature profile is obtained by the resistor network model.

$$T_i = T_c + \frac{\sum_{j=1}^{i} R_j}{\sum_{j=1}^{N} R_j} (T_h - T_c) \tag{2.24}$$

3. According to the corrected temperature profile, the corrected $K_i$'s are obtained.
4. Steps (2) and (3) and repeated until the value of $K_{eff}$ converges.

**2.3. The Coupled System**

The system under consideration is cylindrical. The conduction equation in cylindrical coordinates at steady state (with no internal heat generation) for a fixed thermal conductivity is

$$\frac{1}{r}\frac{d}{dr}\left(r\frac{dT}{dr}\right) = 0 \tag{2.25}$$

Taking the boundary conditions as $T(r_1) = T_c$ (at the inner radius) and $T(r_2) = T_h$ (at the outer radius) and solving the equation yields the cylindrical temperature profile:

$$T(r) = T_h - \frac{T_h - T_c}{\ln(r_1/r_2)} \ln\left(\frac{r}{r_2}\right) \tag{2.26}$$

From the definition of conduction heat transfer,

$$q = -kA\frac{dT}{dr} \tag{2.27}$$

the expression of thermal resistance of a cylindrical system is derived:

$$R = \frac{\ln(r_2/r_1)}{2\pi kL} \tag{2.28}$$

Here, k is the thermal conductivity and L is the length of the cylinder. The curved surface area A in eq. (2.28) is $2\pi rL$.

The resistance is calculated as per eq. (2.28) for each component in the cylindrical system. All calculations are for unit length in this work. To integrate the VDMLI model into this template, an approximate thermal conductivity $k_{vdmli}$ was calculated. For a VDMLI layer within inner and outer radii $r_1$ and $r_2$, with a conductance per unit area $K_{eff}$ and boundary temperatures $T_h$ and $T_c$, its heat transfer in W/m² is

$$q = K_{eff}(T_h - T_c) \tag{2.29}$$

To calculate $k_{vdmli}$, the heat transfer was approximated to be occurring through a curved surface area of $2\pi \frac{r_1+r_2}{2} L$, where $L = 1$. Multiplying this area in eq. (2.29) and equating it with the heat transfer equation written in $q = \Delta T/R$ form, the conductivity is obtained.

$$k_{vdmli} = \left(\frac{r_1 + r_2}{2}\right) \times ln(r_2/r_1) \times K_{eff} \tag{2.30}$$

The corresponding resistance is

$$R_{vdmli} = \frac{1}{\pi K_{eff}(r_1 + r_2)} \tag{2.31}$$

Since the conductance of the VDMLI layer and its temperature profile are dependent on each other, the calculation of boundary temperatures for the VDMLI layer and $K_{eff}$ for the entire VDMLI layer was done through a nested iterative procedure. An in-house Python code was developed to implement the algorithm. Initially, all resistances apart from that of the VDMLI are known. The next steps are described as follows:

1. The temperature profile throughout the entire cylindrical system is assumed. In this work, it is considered as per eq. (2.26).
2. An initial conductance per unit area for the VDMLI is obtained following the process described in section 2.2.
3. The correct boundary temperatures are obtained with the resistor network model (eq. (2.24) applied to the whole system) using this conductance value.
4. These boundary temperatures are fed to the VDMLI loop (in section 2.2) to obtain a corrected temperature profile within the VDMLI and conductance value.
5. Steps (3) and (4) – the two correction stages – are repeated until the conductance value converges. This implies that the two VDMLI boundary temperature values have also converged. The convergence limit is set to 1%.

## 3. Problem Formulation

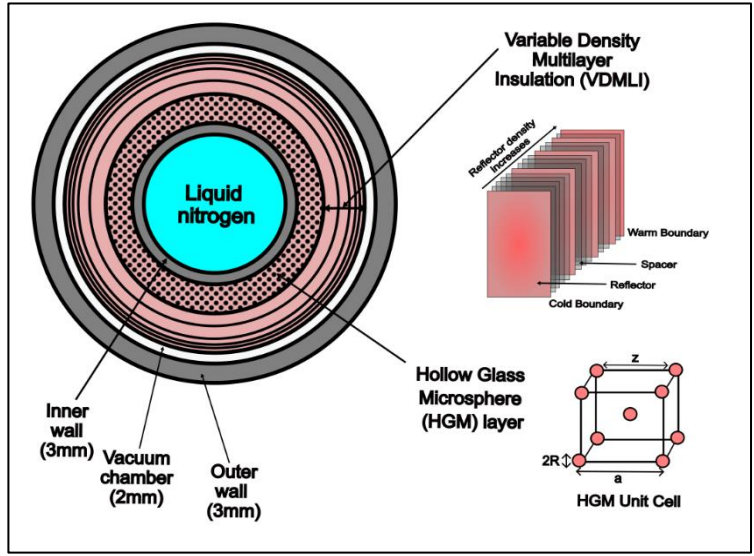

Fig. 1. Schematic of the dewar system cross-section. Representative images illustrate the HGM unit cell and the VDMLI layer.

The cylindrical system considered for the calculations first consists of a 3mm thick inner wall, followed by an HGM layer, a 2mm thick vacuum layer (according to [29]) and a 3mm thick outer wall. Inside the inner wall lies the cryogen (considered to be nitrogen at $T_{\infty 1} = 77K$) and outside the outer wall are ambient conditions at a temperature of $T_{\infty 2} = 303K$. At both the innermost and outermost boundaries of the system, the convection coefficient (eq. (3.4)) is found by evaluating the Nusselt number (eq. (3.3)), which requires the corresponding Prandtl number $Pr$, Grashof number (eq. (3.1)) and Rayleigh number (eq. (3.2)). The term $d$ represents the characteristic dimension of the inner or outer cylinder (i.e. the diameter), and $k$ represents the thermal conductivity of the fluid (liquid nitrogen for the innermost surface, air for the outermost surface) at the given temperature.

$$Gr_d = \frac{g\beta(T_s - T_\infty)d^3}{\nu^2} \tag{3.1}$$

$$Ra_d = Gr_d Pr \tag{3.2}$$

$$Nu = \left[0.6 + \frac{0.387 Ra_d^{1/6}}{(1 + (0.559/Pr)^{1/16})^{8/27}}\right]^2 \tag{3.3}$$

$$h = Nu \times k/d \tag{3.4}$$

The heat transfer per unit length is found by eq. (3.5), where $R_{total}$ represents the total thermal resistance of the system.

$$q = (T_{\infty 1} - T_{\infty 2})/R_{total} \tag{3.5}$$

For the second part of the paper, a layer of VDMLI was added just after the HGM layer to explore their combined insulation effect and assess the performance of the VDMLI in the presence of an HGM layer of optimal thickness and no pressure leaks. The vacuum pressures in the HGM layer and the VDMLI layer are controlled independently.

## 4. <u>Validation</u>

In Fig. 2(a), the values of thermal conductivity obtained by implementing the HGM Layer model is compared with experimental values given in Fig. 6 of the ASTM Standard Guide 2019 [31]. The thermal conductivity values approximately match in the range of 4 Pa to 133 Pa (30 millitorr to 1000 millitorr). For lower pressures, the effect of the residual gas in the HGM layer becomes practically negligible. This may be attributed to the movement of residual gas molecules practically ceasing at very low pressures, making conductivity assume a nearly constant value. On the other hand, the analytical model always considers the effect of residual gas molecules in conductivity calculation and thus fails to attain the experimental values to the left of approximately 10Pa (75 millitorr) pressure.

In Fig. 2(b), test cases with the algorithm described in section 2.2 are shown. A simple system, consisting only of a 100mm-thick VDMLI layer with boundary temperatures 78K and 293K, is considered for different tank sizes. Four arrangements are simulated: two arithmetic progression-based sequences of reflectors (represented by slashes) and spacers (represented by numbers) and their reverse cases. 1/2/3/… (A1) means that the sequence starts on the cold side with 1 spacer followed by 1 reflector, 2 spacers, 1 reflector and so on. Sequences A1 and A2 are sequences wherein the reflector density is higher on the cold side; in their reverse sequences, the reflector density is higher on the hotter side. It is seen that the conductance is significantly lesser for the reverse sequences, consistent with the trends in optimal arrangements reported in B Wang et al. [21]. The reasoning that a higher number of reflectors on the hotter side is significantly more effective in blocking radiation holds for these test cases. The graph also suggests that the VDMLI conductance, and the results associated with it, are invariant across tank sizes – thus, in section (5.2), results for one of the two tank sizes considered in section (5.1) are analysed.

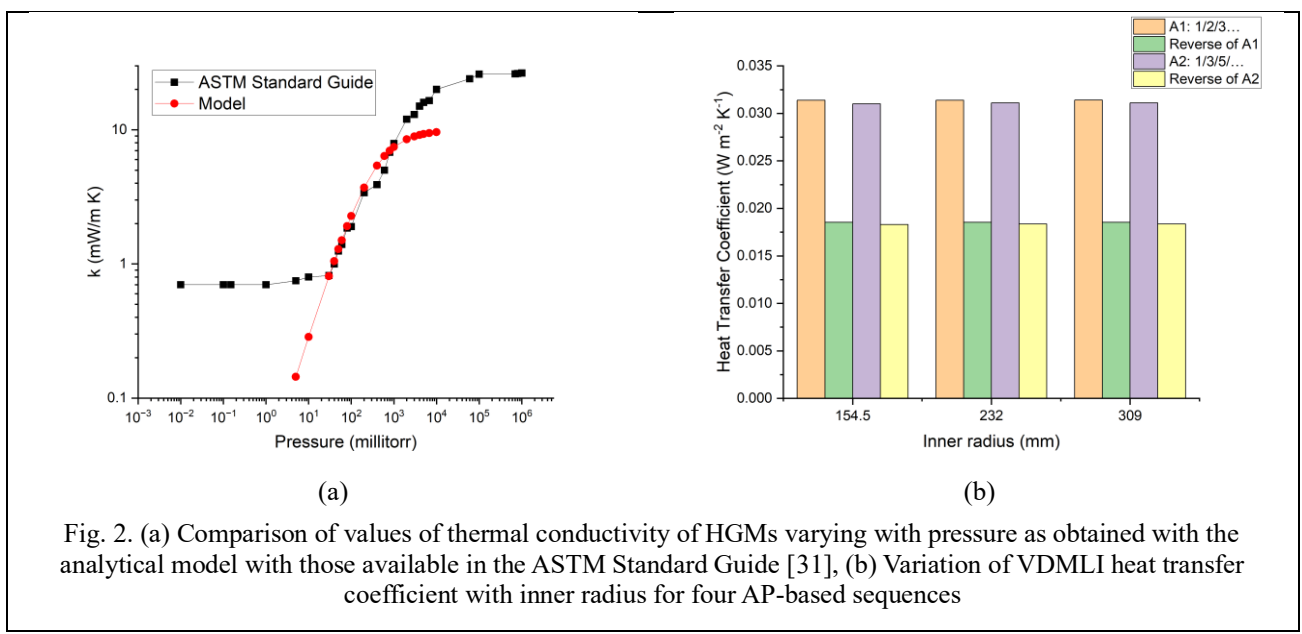

Fig. 2. (a) Comparison of values of thermal conductivity of HGMs varying with pressure as obtained with the analytical model with those available in the ASTM Standard Guide [31], (b) Variation of VDMLI heat transfer coefficient with inner radius for four AP-based sequences

# 5. Results

## 5.1. HGM Layer Thickness Optimization

Figures (3) and (4) depict the variation in heat transfer through the HGM layer with its thickness. They are based on five reference vacuum pressures between 0.1 and 750000 millitorr (i.e. between high vacuum and atmospheric pressure regions) and inner cylinder diameters (ID) of 145mm [30] and 309mm [29]. The conductivity for HGMs beyond the limits of the pressure range satisfied by the analytical model is taken from the experimental results as reported in [31]. It is observed that after a certain thickness, the variation in heat transfer becomes negligibly small. This thickness is concluded to be optimal for effective insulation performance of the cryogenic storage system. The cryogen is taken as liquid nitrogen at 77K.

The HGM thickness at pressures of 0.1 and 75 millitorrs (Fig. 2), in case of both the diameters, appears to stabilize with respect to heat transfer values at approximately 40% (140mm compared to 350mm in Fig. 2) and 37.5% (90mm compared to 240mm in Fig. 3) of the HGM layer thickness at higher pressure. From the literature [29, 30], it is seen that the ideal pressure region to operate the insulation is high vacuum, followed by soft vacuum. Further, the effect of HGM thickness on heat transfer is next seen at intermediate pressures between 750Pa and atmospheric pressure ($10^5$ Pa) for the reference dimensions of the system. This is done for the purpose of predicting the ideal thickness that could be used in case there is a deviation in ideal pressure conditions, such as a vacuum leak. For ID 309mm (Fig. (2)), the optimal thickness is around 160mm for 750Pa while for both the higher-pressure values, it is around 225mm. For ID 145mm (Fig. (3)), which displays a similar trend, the optimal thickness for all the reference pressures is observed to be around 130mm.

From a collective observation of Fig. (3) and Fig. (4), it appears that a total HGM layer thickness between 200mm and 250mm, with the lower limit being about 10-15% higher in case of higher vacuum pressure conditions, is optimal for insulation in case of liquid nitrogen. This holds true especially if the system is susceptible to vacuum leak. However, on closer analysis with percentage change in heat transfer (Fig. (5) and Fig. (6)), it is revealed that the decrease in heat transfer with respect to its immediate previous value at a given thickness almost saturates around 120-130mm for both the vessel sizes. For the smaller diameter (Fig. (6)), the percentage decrease falls to around 5% from 17%, and for the larger diameter (Fig. (5)) it falls to around 10% from 20%.

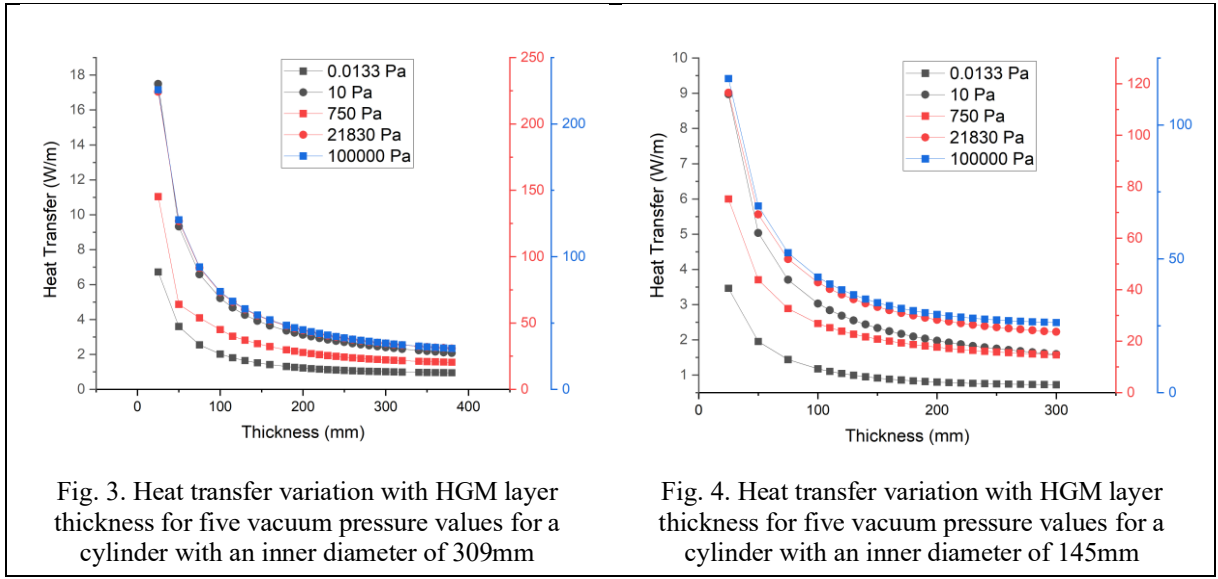

Fig. 3. Heat transfer variation with HGM layer thickness for five vacuum pressure values for a cylinder with an inner diameter of 309mm

Fig. 4. Heat transfer variation with HGM layer thickness for five vacuum pressure values for a cylinder with an inner diameter of 145mm

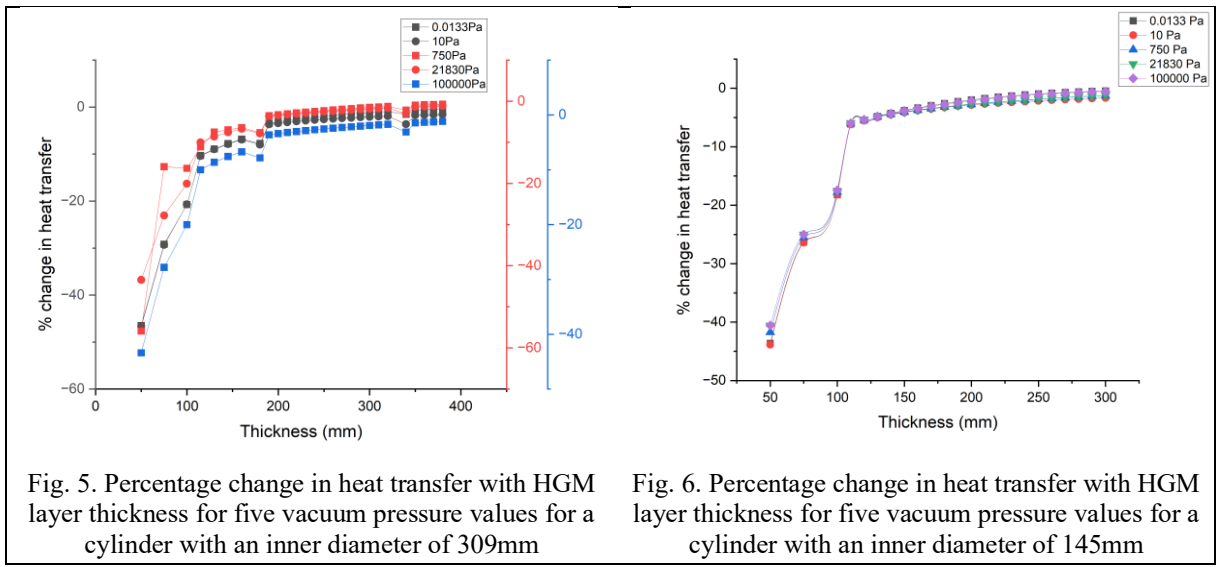

Fig. 5. Percentage change in heat transfer with HGM layer thickness for five vacuum pressure values for a cylinder with an inner diameter of 309mm

Fig. 6. Percentage change in heat transfer with HGM layer thickness for five vacuum pressure values for a cylinder with an inner diameter of 145mm

### 5.2. VDMLI Optimization in a Coupled System

In the second part of the problem, a VDMLI layer was added to the insulation system just after an optimal HGM layer. According to the results in section (5.1), the optimal thickness of the HGM layer is taken as 130mm and it is maintained at 0.0133 Pa vacuum pressure for all cases. 0.5-mil aluminized reflectors and 0.00635mm dacron net spacers were considered for the VDMLI. The simulations were performed for one of the cryogenic vessels taken in section (5.1) whose inner diameter is 309mm.

The VDMLI sequences considered were arithmetic progression (AP)-based. In the linearly increasing arrangement, the number of spacers in every layer was incremented. Two such arrangements were considered: 1/2/3/… (common difference 1) and 1/3/5… (common difference 2). In the step-increasing arrangement, every spacer layer was repeated once. Two such arrangements were considered, in synchronization with their linear counterparts: 1/1/2/2/… (common difference 1) and 1/1/3/3/… (common difference 2). In consistency with the trend observed in Fig. 1(b), in all simulations the high-density VDMLI (i.e. greater number of reflectors) is taken on the hotter side.

In Fig. (7), the variation of the conductance per unit area for VDMLI (or the VDMLI heat transfer coefficient) with vacuum pressure is shown. The thickness is 100mm with a tolerance of 0.8%, which means that the nearest thickness value greater or less than 100mm that can be achieved with a given arrangement is considered. It is observed that for high vacuum pressures, the linear arrangements (values on the left-hand-side Y-axis) perform slightly better the step arrangements (values on the right-hand-side Y-axis) whose conductance is around 0.01 W m$^{-2}$ K$^{-1}$. However, this behaviour flips at a higher pressure (around $10^4$ Pa), far beyond the vacuum pressure range. It is seen that at atmospheric pressure, the conductance of the 1/1/2/2/… arrangement is nearly 18% less than the highest conductance value at that pressure (for 1/3/5/7/…). Comparison between these two arrangements also show, in contrast, that at high vacuum (0.001 Pa) the 1/3/5/7/… arrangement performs better by 7%. Additionally, flips in the VDMLI behaviour are also observed individually in the linear and step arrangements: after 10 Pa, the 1/2/3/4/… arrangement behaves better than the 1/3/5/7/… arrangement, and after 50 Pa the 1/1/2/2/… arrangement behaves better than the 1/1/3/3/… arrangement. These changes can be attributed to a lesser density of reflectors with increasing thickness in the 1/3/5/7/… and 1/1/3/3/… arrangements with respect to their unity common difference counterparts. However, as far as optimality is concerned, difference in insulation between the linear and step arrangements is much more significant at higher pressures. In case of pressure leaks, the step arrangements (and specifically the 1/1/2/2/… arrangement) perform significantly better than linear ones in presence of an optimal HGM layer.

To further characterize the VDMLI performance, the performance of the optimal arrangement 1/1/2/2/… is visualized for higher warm boundary temperatures up to 393K. Fig. 8(a) shows the VDMLI conductance per unit area with boundary temperatures ranging from 293K to 393K in increments of 20K for six thicknesses. The value of heat transfer at a given warm boundary temperature decreases with increasing thickness, as is expected; in addition, the slope of the linear variations decreases with increasing thickness. The conductance at 125mm appears to be optimal, with 0.06 W m$^{-2}$ K$^{-1}$ at 393K (the most extreme condition) being 53% lower than the conductance for 100mm and only 17% higher than the conductance for 150mm. The corresponding differences are 33% and 17% for 293K. The difference in conductance at higher warm boundary temperatures decreases further down the graph as thickness increases. Furthermore, it is observed that at any warm boundary temperature, the change in heat transfer through the system is very minimal (in the order of $10^{-3}$ W/m) with varying thickness. The heat transfer values averaged over all six thicknesses for the considered warm boundary temperatures are reported in Fig. 8(b). This implies that in comparison to the optimal 130mm-thick HGM at high vacuum conditions, the VDMLI underperforms significantly. At best, however, the VDMLI can be used as a safety measure against vessel leaks if an HGM layer more than 100mm thick is used. Further combinations may be obtained to optimize the cost of the insulation system as per requirement.

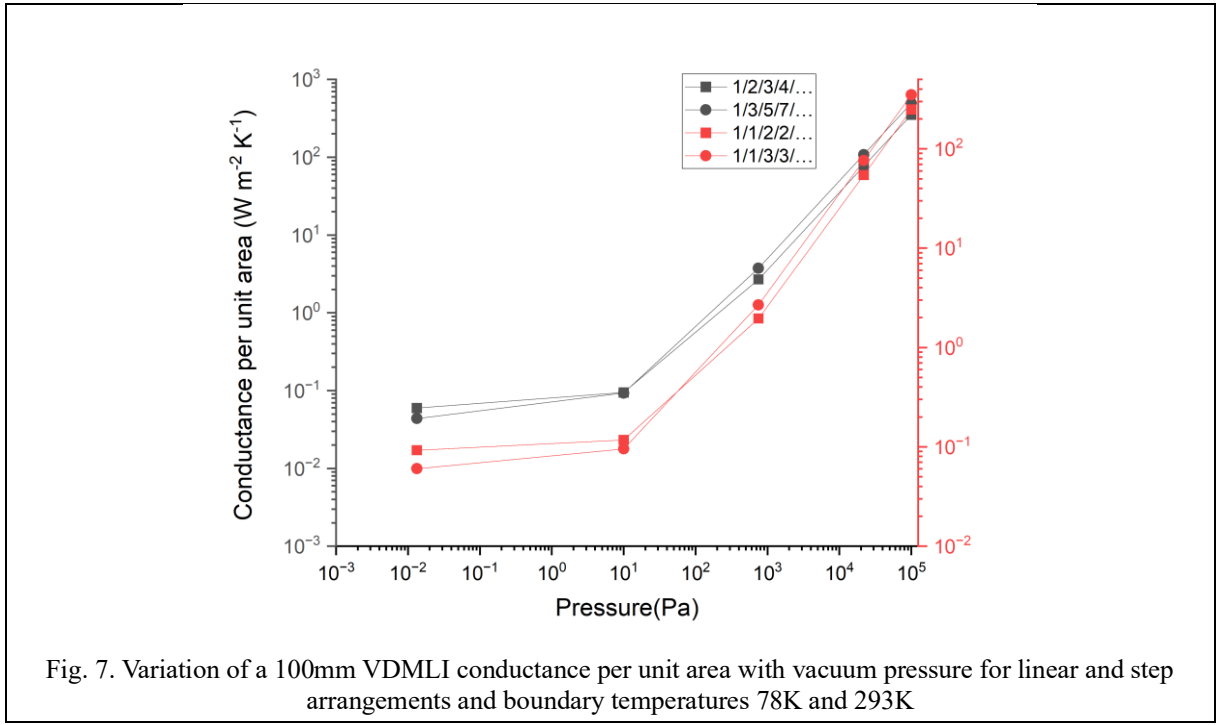

Fig. 7. Variation of a 100mm VDMLI conductance per unit area with vacuum pressure for linear and step arrangements and boundary temperatures 78K and 293K

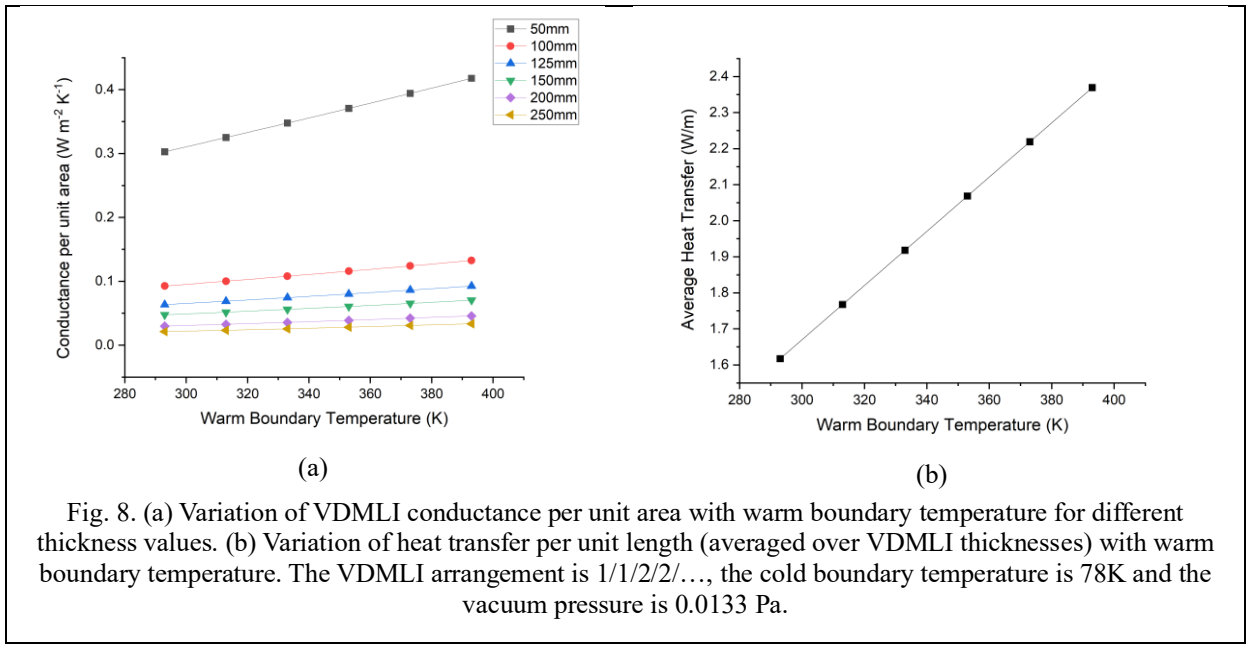

(a)                                              (b)

Fig. 8. (a) Variation of VDMLI conductance per unit area with warm boundary temperature for different thickness values. (b) Variation of heat transfer per unit length (averaged over VDMLI thicknesses) with warm boundary temperature. The VDMLI arrangement is 1/1/2/2/…, the cold boundary temperature is 78K and the vacuum pressure is 0.0133 Pa.

## 6. Conclusion

The study analyses the performance of a composite insulation system of HGMs and VDMLIs. First, the HGM layer was modelled analytically based on a unit cell and validated for a vacuum pressure range of 4 Pa to 133 Pa (soft vacuum). Next, the heat transfer through a cylindrical insulation system containing an HGM layer was simulated for two cryogenic vessels of inner diameters 309mm and 145mm for five pressures ranging from 0.0133 Pa (high vacuum) to atmospheric pressure (pressure leak condition). In the resistor network model for the system, convection from the cryogen to the inner wall and from the outer wall to ambient were also considered. Then, the VDMLI layer was modelled using the existing OVDMLI and the layer-by-layer models and an iterative procedure was implemented to find the conductance per unit area. The VDMLI was introduced into the insulation system for the cryogenic vessel of inner diameter 309mm in combination with an optimal HGM thickness of 130mm at 0.0133 Pa pressure. The variation of VDMLI conductance with pressure (in the same range used earlier) was reported for linear and step arrangements for VDMLI. The most optimal arrangement was chosen, and its conductance was used to characterize the VDMLI performance and heat transfer through the system by increasing the warm boundary temperature up to 393 K. The key findings are summarized as follows:

- A range of thickness 120-130mm was reported as optimal for the HGM layer for both sizes of the cryogenic cylinder, the change in heat transfer being less than 5% for the smaller vessel and 10% for the larger vessel beyond this thickness.
- In combination with a VDMLI layer, it was seen that a 1/1/2/2/… arrangement of reflectors and spacers performs the best among simulated linear and step arrangements in pressure leak conditions.
- The optimal thickness for the 1/1/2/2/… arrangement was reported to be 125mm, the decrease in conductance changing from 33-53% to below 17% hereafter with increasing thickness for warm boundary temperatures 293 K to 393 K.
- In high vacuum, the HGM outperforms the VDMLI significantly and when both are used in the insulation system, heat transfer per unit length is negligibly affected. Therefore, based on its performance, an optimal VDMLI can be useful as a protective layer on an optimal HGM at best.

However, optimizing the composite cryogenic insulation system comprising of HGMs and VDMLIs is a process dependent on several factors: the HGM and VDMLI's individual insulation performance, their combined effect, their weight, the costs, the space available et cetera. The combined optimal HGM-VDMLI thickness for practical situations is subject to some of or all these constraints, and there is scope to characterize individual and combined optimal thicknesses based on specific applications. Further studies are possible in VDMLI optimization algorithms, a process that may benefit from application of advanced computational tools like machine learning, and analytical models may also expand on various factors such as the conductivity variation of HGM assemblies and VDMLIs with temperature and convective effects inside the chamber which will allow for more precise calculations.

**Declaration of Competing Interests**

The authors report there are no competing interests to declare.

# List of Notations

| Notation | Meaning |
|---|---|
| $\eta_0$ | Initial packing fraction |
| $\eta$ | Final packing fraction |
| $N_o$ | Overlap coordination number |
| $N_g$ | Gap coordination number |
| $C_o^e$ | Overlap conductance |
| $C_g^e$ | Gap conductance |
| $D$ | Diameter of hollow glass microspheres (HGMs) |
| $R$ | Radius of hollow glass microspheres (HGMs) |
| $\epsilon_r$ | Emissivity (other subscripts: h – hot, c – cold) |
| $S$ | Stefan-Boltzmann constant |
| $k_r$ | Radiative thermal conductivity |
| $k_{eff}$ | Effective thermal conductivity |
| $z$ | Maximum spacing between two microspheres |
| $a$ | Unit cell side length |
| $\zeta$ | Fitting parameter |
| $\xi$ | Dimensionless gap conductance |
| $\gamma$ | Efficiency of energy transfer between gas and microsphere |
| $a_c$ | Accommodation coefficient |
| $K_n$ | Knudsen number |
| $\lambda$ | Mean free path |
| $L_g$ | Characteristic dimension of gap |
| $K_f^c$ | Gas conductivity at contact between microspheres |
| $T$ | Mean of warm and cold boundary temperatures |
| $P$ | Vacuum pressure |
| $m_f$ | Molar mass of gas |
| $R_g$ | Universal gas constant |
| $g$ | Acceleration due to gravity |
| $\beta$ | Coefficient of volume expansion |
| $T_s$ | Temperature at cylinder surface |
| $K_{eff}$ | Effective conductance per unit area (heat transfer coefficient for VDMLI) |
| $h$ | Convective heat transfer coefficient |
| $Gr_d$ | Grashof number |
| $Pr$ | Prandtl number |
| $Ra_d$ | Rayleigh number |
| $Nu$ | Nusselt number |